\documentclass[twocolumn,prl,floatfix,superscriptaddress,nofootinbib]{revtex4-1}

\usepackage{graphicx}
\usepackage{amsmath,amssymb}
\usepackage{color}

\graphicspath{{./Figures/}}

\begin{document}

\title{All-optical sensing of a single-molecule electron spin}

\author{A. O. Sushkov}
\thanks{These authors contributed equally to this work}
\affiliation{Department of Physics, Harvard University, Cambridge, MA 02138, USA}
\affiliation{Department of Chemistry and Chemical Biology, Harvard University, Cambridge, MA 02138, USA}
\author{N. Chisholm}
\thanks{These authors contributed equally to this work}
\affiliation{School of Engineering and Applied Sciences, Harvard University, Cambridge, MA 02138, USA}
\author{I. Lovchinsky}
\thanks{These authors contributed equally to this work}
\affiliation{Department of Physics, Harvard University, Cambridge, MA 02138, USA}
\author{M. Kubo}
\affiliation{Department of Chemistry and Chemical Biology, Harvard University, Cambridge, MA 02138, USA}
\author{P. K. Lo}
\affiliation{Department of Biology and Chemistry, City University of Hong Kong, Tat Chee Avenue, Kowloon, Hong Kong SAR, China}
\author{S. D. Bennett}
\affiliation{Department of Physics, Harvard University, Cambridge, MA 02138, USA}
\author{D. Hunger}
\affiliation{Max-Planck-Institut f\"{u}r Quantenoptik,  Garching, D-85748, Germany}
\author{A. Akimov}
\affiliation{Russian Quantum Center, Skolkovo, Moscow Region, 143025 Russia}
\author{R. L. Walsworth}
\affiliation{Department of Physics, Harvard University, Cambridge, MA 02138, USA}
\affiliation{Harvard-Smithsonian Center for Astrophysics, Cambridge, Massachusetts 02138, USA}
\affiliation{Center for Brain Science, Harvard University, Cambridge, Massachusetts 02138, USA}
\author{H. Park}
\email{Hongkun\textunderscore Park@harvard.edu}
\affiliation{Department of Physics, Harvard University, Cambridge, MA 02138, USA}
\affiliation{Department of Chemistry and Chemical Biology, Harvard University, Cambridge, MA 02138, USA}
\affiliation{Broad Institute of MIT and Harvard, 7 Cambridge Center, Cambridge, MA, 02142, USA}
\author{M. D. Lukin}
\email{lukin@physics.harvard.edu; corresponding author}
\affiliation{Department of Physics, Harvard University, Cambridge, MA 02138, USA}

\begin{abstract}
We demonstrate an all-optical method for magnetic sensing of individual molecules in ambient conditions at room temperature. Our approach is based on shallow nitrogen-vacancy (NV) centers near the surface of a diamond crystal, which we use to detect single paramagnetic molecules covalently attached to the diamond surface. The manipulation and readout of the NV centers is all-optical and provides a sensitive probe of the magnetic field fluctuations stemming from the dynamics of the electronic spins of the attached molecules. As a specific example, we demonstrate detection of a single paramagnetic molecule containing a gadolinium (Gd$^{3+}$) ion. We confirm single-molecule resolution using optical fluorescence and atomic force microscopy to co-localize one NV center and one Gd$^{3+}$-containing molecule. Possible applications include nanoscale and \textit{in vivo} magnetic spectroscopy and imaging of individual molecules.
\end{abstract}

\maketitle

Precision magnetic sensing is essential to a wide array of technologies such as magnetic resonance imaging (MRI), with important applications in both the physical and life sciences. In particular, in biology and medicine, functional magnetic resonance imaging (fMRI) has emerged as a primary workhorse for obtaining key physiological and pathological information noninvasively, such as blood and tissue oxygen level and redox status~\cite{Logothetis2008,Matsumoto2007,Ahrens2013}. Developing nanoscale magnetic sensing applicable to individual molecules could enable revolutionary advances in the physical, biological, and medical sciences. Examples include determining the structure of single proteins and other biomolecules as well as \textit{in vivo} measurements of small concentrations of reactive oxygen species that could lead to insights into cellular signaling, ageing, mutations, and death~\cite{Netzer2009,Halliwell2007,Valko2007,James2004}. The practical realization of these ideas is extremely challenging, however, as it requires sensitive detection of weak magnetic fields associated with individual electronic or nuclear spins at nanometer scale resolution, often under ambient, room-temperature conditions.
Many state-of-the-art magnetic sensors, including superconducting quantum interference devices (SQUIDs)~\cite{Nowack2013}, semiconductor Hall effect sensors~\cite{Bending1999}, and spin exchange relaxation-free atomic magnetometers~\cite{Allred2002a}, offer outstanding sensitivity, but their macroscopic nature precludes individual spin sensing. Sensing ensembles of paramagnetic molecules in biological and medical systems is currently performed using bulk electron spin resonance (ESR), which has a limit of roughly $10^{7}$ electron-spins~\cite{Blank2004}. Magnetic resonance force microscopy has been used to detect individual electronic spins, but operates at cryogenic, milliKelvin temperature~\cite{Rugar2004,Degen2009}.

In this Letter we demonstrate a method for nanoscale magnetic sensing of individual non-fluorescent molecules that employs optical manipulation of nitrogen vacancy (NV) centers in diamond~\cite{Taylor2008,Maze2008,Balasubramanian2008,Neumann2010,Steinert2013,Kaufmann2013,Staudacher2013,Mamin2013}.
In our approach the target molecules are covalently attached to the diamond surface, and magnetic sensing of these molecules is performed under ambient conditions using a single shallow NV center as an all-optical nanoscale magnetometer (Fig. 1A). Importantly, the shallow NV center is close enough to the surface that it can detect the fluctuating magnetic field produced by the electronic spin of a single molecule, while maintaining good NV center spin coherence and optical properties. We apply this technique to detect a single paramagnetic molecule composed of a gadolinium ion (Gd$^{3+}$) chelated by an amine-terminated organic ligand (abbreviated as Gd$^{3+}$ molecule below). Single-molecule sensing is confirmed by identifying NV centers that have only a single target molecule within the sensing area on the diamond surface.
\begin{figure}[h]
 %\begin{center}
 \includegraphics[width=\columnwidth]{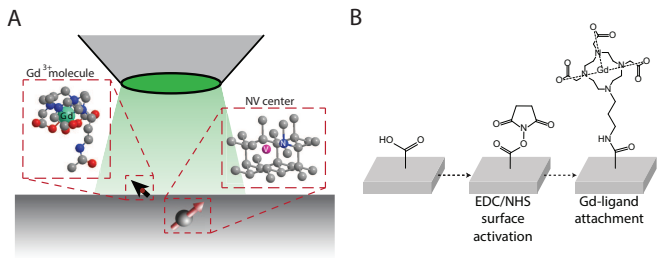}
 %\end{center}
 \caption{Schematic of measurement setup and sample preparation.
(A) Single-molecule electron spin detection using a single shallow NV center in bulk diamond. Gd$^{3+}$ molecules are attached to the surface of a bulk diamond crystal with widely separated NV centers located at a nominal depth of 6 nm below the diamond surface. NV center optical pumping and fluorescence detection is performed using a confocal microscope (objective shown).
(B) Chemical procedure for attaching Gd$^{3+}$ molecules to the diamond surface. 1-ethyl-3-(3-dimethylaminopropyl)carbodiimide (EDC) and \textit{N}-hydroxysulfosuccinimide (NHS) are used to activate carboxyl groups on the diamond surface so that they react readily with Gd$^{3+}$ molecules functionalized with amine groups.
}
 \label{fig:Figure1}
 \end{figure}

Our scheme for covalently attaching molecules to the diamond surface relies on the coupling of the amine-functionalized Gd$^{3+}$ molecule to the carboxylic group on the diamond surface: in order to improve this coupling efficiency, we activated the surface carboxylic group using a water-soluble mixture of 1-ethyl-3-(3-dimethylaminopropyl)carbodiimide (EDC) and \textit{N}-hydroxysulfosuccinimide (NHS) (Fig. 1B). This method yielded  uniform surface coverage of molecules, with little clumping~\cite{som}, and the surface density of molecules could be controlled by varying the concentration of the Gd$^{3+}$ molecules during the reaction. This procedure can be used to covalently attach any water-soluble amine-terminated molecule to the diamond surface, with controlled surface coverage.  Since covalent attachment utilizes diamond surface carboxylic groups, the resulting molecular surface density was always less than a monolayer.

In our experiments, we used atomic force microscopy (AFM) to quantify the surface density of these molecules, and to identify their proximity to a given shallow NV center. AFM measurements show that a clean diamond surface exhibits atomically-smooth regions of typically a few square micrometers. When the Gd$^{3+}$ molecules were attached, we observed circular features with mean height of 8~$\AA$ in the AFM scans. The heights, radii, and density of these features were consistent with single Gd$^{3+}$ molecules covalently attached to the diamond surface~\cite{som}. As an independent check of the surface molecule density, we added  a single Cy3 dye molecule to each Gd$^{3+}$ molecule, and then attached the resulting molecule to the diamond surface using the same chemical procedure as before. We then performed surface fluorescence measurements to deduce the Cy3 surface density, and compared the result to the density of the surface molecules measured using AFM. The results of these independent measurements were consistent with each other, providing strong evidence that the 8 $\AA$-high AFM features are indeed single molecules~\cite{som}.

 \begin{figure}[b]
 %\begin{center}
 \includegraphics[width=1.0\columnwidth]{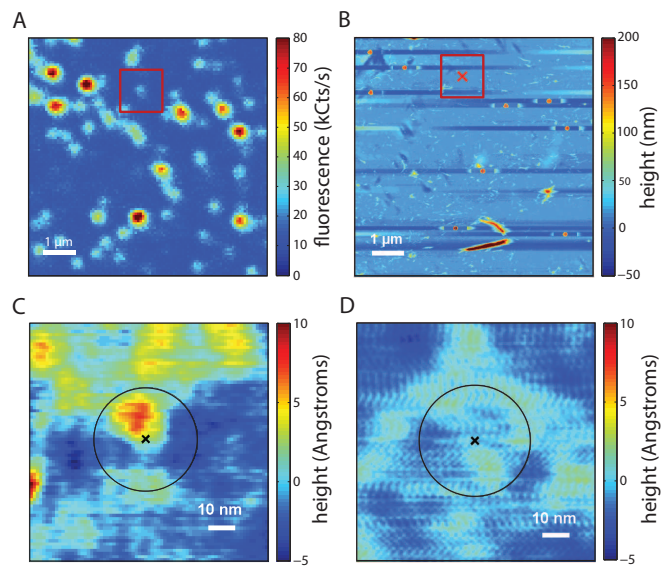}
 %\end{center}
 \caption{Co-localization of a single shallow NV center and a Gd$^{3+}$ molecule.
(A) Fluorescence image of a 7.5 $\mu$m $\times$ 7.5 $\mu$m area of the diamond crystal, showing several gold nanoparticles (bright spots), and NV centers (less intense spots). Location of a single NV center, marked by a red square, was determined in relation to the gold nanoparticles.
(B) AFM image of the same region of the diamond surface, showing gold nanoparticles (red dots). The red cross marks the location of the NV center, deduced from the fluorescence image.
(C) AFM image of the 100 nm $\times$ 100 nm region of the diamond surface centered at the location of the NV center (marked by a black cross). The black circle shows the one standard deviation uncertainty in the NV center position, with a single Gd$^{3+}$ molecule present within the circle (bright spot).
(D) AFM image of the same area as in (C), after Gd$^{3+}$ molecules were removed from the diamond surface.}
 \label{fig:Figure2}
 \end{figure}
In order to determine the proximity of Gd$^{3+}$ molecules to a given shallow NV center with nanoscale precision, we performed a three-step co-localization experiment (Fig. 2). First, we coated the diamond surface (via electrostatic attachment) with 100 nm-diameter gold nanoparticles that fluoresce in the same spectral region as the NV centers and are optically resolvable individually. Second, we performed a fluorescence scan to determine the locations of individual NV centers and gold nanoparticles optically (Fig. 2A). Finally, we performed AFM topography measurements to determine the locations of gold  nanoparticles and Gd$^{3+}$ molecules (Fig. 2B). Because the nanoparticles appear in both optical and AFM images, we can use the locations of nanoparticles to combine the fluorescence and AFM measurements and deduce the lateral positions of Gd$^{3+}$ molecules relative to an NV center, with uncertainty of approximately 15 nm~\cite{som}. Figure 2C shows an example of this co-localization experiment: an AFM image of Gd$^{3+}$ molecules together with the position of a single shallow NV center, marked by a cross (the circle shows the NV center position uncertainty at one standard deviation). When the Gd$^{3+}$ molecules were removed from the diamond surface, the AFM scan of the same region showed the absence of 8 $\AA$-high features, confirming the successful removal of molecules (Fig. 2D).

 \begin{figure}[!b]
 %\begin{center}
 \includegraphics[width=\columnwidth]{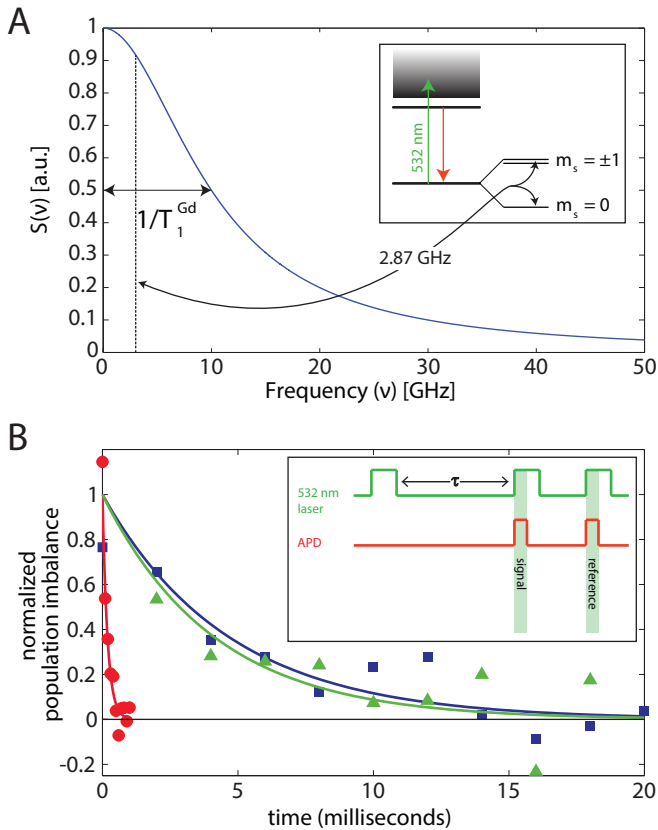}
 %\end{center}
 \caption{Measurement of magnetic noise from a single Gd$^{3+}$ molecule attached to a diamond surface using a single shallow NV center.
(A) Schematic power spectrum of the fluctuating magnetic field due to relaxation of the Gd$^{3+}$ electronic spin (inset: NV center electronic excited and ground states, with ground-state spin sublevels). Fourier components of this spectrum near the frequency resonant with the NV center zero-magnetic-field splitting lead to an increase in the NV center spin-state population relaxation rate.
(B) Demonstration of NV magnetic sensing of a single Gd$^{3+}$ molecule on the surface of bulk diamond. Measurements of the NV center spin-state population difference relaxation and exponential fits. Clean diamond surface: blue squares and blue line. Gd$^{3+}$ molecules attached to the diamond surface: red circles and red line. Re-cleaned diamond surface: green triangles and green line. The AFM image for this NV center is shown in Fig. 2C, where it is demonstrated that it is in proximity to a single Gd$^{3+}$ molecule. The scatter of the experimental data points is consistent with photon shot noise, with total averaging time on the order of a few hours.
Inset: Pulse measurement scheme for measuring the NV center spin-state relaxation rate. An avalanche photodiode (APD) was used for NV center red fluorescence detection.
}
 \label{fig:Figure3}
 \end{figure}
Once we located a single Gd$^{3+}$ molecule with a nearby NV center, we performed all-optical magnetic sensing of this molecule. At room temperature the $S = 7/2$ electron spin of the Gd$^{3+}$ ion fluctuates with a relaxation rate ($\gamma_{Gd}$) on the order of 10~GHz~\cite{Bierig1964,Kim2009b}. These spin-flips give rise to a fluctuating magnetic field at the location of the NV center, with a Fourier spectrum of width $\approx \gamma_{Gd}$. The Fourier component of this fluctuating magnetic field at the frequency corresponding to the zero-field splitting of the NV center ground state spin manifold (S = 1) drives magnetic dipole transitions between these sublevels (Fig. 3A). We detected these transitions by first optically pumping the NV center into the $m_{s} = 0$ sublevel, and then measuring its spin-state-dependent fluorescence after a variable delay time $\tau$ (Fig. 3B, inset). In the absence of Gd$^{3+}$ molecules, the NV spin-state population difference decayed with rate $\Gamma_{intrinsic}$ due to spin-lattice relaxation. However, when the NV center was in proximity to a Gd$^{3+}$ molecule, the measured NV population relaxation rate increased to $\Gamma_{total} = \Gamma_{intrinsic} + \Gamma_{induced}$~\cite{som}, which constitutes magnetic sensing of single-molecule electron spin. For example, the red circles in Fig. 3B show the result of the NV spin-state relaxation measurements for the NV-Gd$^{3+}$ molecule pair displayed in Fig. 2C (the red line is an exponential fit); while the blue squares in Fig. 3B illustrate the spin-state relaxation rate of the same NV center prior to attachment of the Gd$^{3+}$ molecule. The comparison of these measurements clearly shows a dramatic increase of the relaxation rate in presence of a single Gd$^{3+}$ molecule. Once the molecule was removed (Fig. 2D), the relaxation returned to the intrinsic rate (green triangles in Fig. 3B).

 \begin{figure}
 %\begin{center}
 \includegraphics[width=\columnwidth]{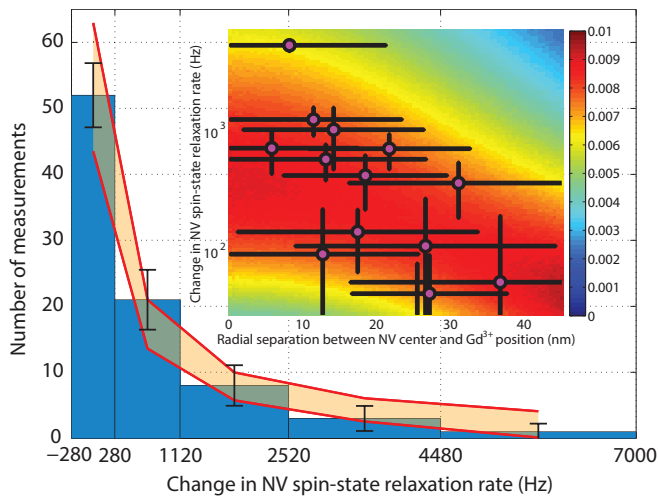}
 %\end{center}
 \caption{Magnetic noise measurements in the single Gd$^{3+}$ molecule sensing regime.
Results of 85 Gd-induced NV center spin-state relaxation rate measurements, along with a Monte-Carlo simulation band. The experimental data are grouped into five bins, with the error bars calculated by combining bin sampling uncertainty and relaxation rate fitting uncertainty~\cite{som}. The theoretical band was obtained from Monte Carlo simulations of the experiment, with  parameters given in the text.
Inset: results of 14 co-localization and NV center spin-state relaxation rate measurements in which a single Gd$^{3+}$ molecule was identified near a single shallow NV center. The background displays the results of Monte Carlo simulations of the experiment, with the color scale indicating the probability density of obtaining a particular NV center spin-state relaxation rate for a given separation between the NV center and the proximal Gd$^{3+}$ molecule. The simulation was performed for 20~nm separation between Gd$^{3+}$ molecules, and NV center co-localization uncertainty of 15~nm. For a quantitative comparison, we performed a two-variable Kolmogorov-Smirnov statistical test, resulting in the Z-statistic value of 1.1, which indicates that the data points are consistent with the simulated distribution~\cite{som}.
}
 \label{fig:Figure4}
 \end{figure}
The inset of Figure 4 summarizes the measured Gd-induced relaxation rates of for several NV-Gd$^{3+}$ molecule pairs, with varying NV-molecule separations. We performed a total of 23 co-localization experiments, together with population relaxation measurements of the corresponding NV centers. In 14 of the 23 co-localization experiments, we could reliably identify single Gd$^{3+}$ molecules and extract the separation between an NV center and a Gd$^{3+}$ molecule; while in the remaining 9 experiments, we could not do so because of finite AFM tip resolution or rough surface topography. Seven of the data points exhibit a significant (greater than two standard deviation) increase in NV spin relaxation, and the corresponding co-localization measurements show the presence of a single Gd$^{3+}$ molecule near the NV center position.  As noted above, removal of the Gd$^{3+}$ molecules from the diamond surface resulted in the relaxation rate returning to its intrinsic value in all cases.

A comparison of these data with Monte Carlo simulations (shown as background color plot in the inset of Fig. 4) provides further evidence of NV magnetic detection of a single-molecule electron spin. In the simulation, we calculated the probability density of obtaining a particular NV spin relaxation rate for a given NV-Gd$^{3+}$ molecule separation~\cite{som} within experimental uncertainties. We used an NV center depth of 6 nm, derived from calculations for 3 keV nitrogen ion implantation energy; a mean Gd$^{3+}$ molecule spacing of 20 nm, derived from the AFM and Cy3 measurements described above; and a Gd$^{3+}$ spin-relaxation rate of  10 GHz~\cite{Bierig1964,Kim2009b,som}. As seen in the inset of Fig. 4, the experimental data points are consistent with the simulated probabilities~\cite{som}.

Additional evidence for magnetic detection of single-molecule electron spins is provided by an independent set of 85 spin relaxation rate measurements that we performed on 26 shallow NV-centers over several cycles of Gd$^{3+}$ molecule attachment and removal. As shown in the main plot of Fig. 4, the resulting data are grouped into five bins, with the error bars calculated by combining the bin sampling and relaxation rate fitting uncertainties~\cite{som}. Also shown in this figure is a band of theoretically calculated NV spin relaxation rates, which we obtained from Monte Carlo simulations of the experiment, with the NV center depth of 6~nm, Gd$^{3+}$ spin-flip rate varying in the range 10 to 20 GHz, and mean Gd$^{3+}$ surface density varying in the range 1/(20 nm)$^2$ to 1/(25 nm)$^2$. These parameters yield simulated NV-center spin relaxation rate distributions that are consistent with experimental data, again confirming that the observed NV spin relaxation  rate increase is due to the proximity of a single molecule electron spin.
While other sets of model parameters can, in principle, be fit to the experimental data, all realistic model fit parameters correspond to regimes in which only a single Gd$^{3+}$ spin contributes to increased NV center spin-state relaxation rate~\cite{som}.
The ``sensing radius'' of an NV-center (defined as the NV-Gd$^{3+}$ molecule separation for which the change in NV-center spin relaxation rate is equal to the measurement uncertainty) is determined to be approximately 12~nm.
This means that, with probability over 80\%, only a single Gd$^{3+}$ molecule can substantially contribute to the induced NV-center spin relaxation rate even for highest Gd$^{3+}$ molecule densities used.

Our method for all-optical magnetic sensing of single paramagnetic molecules using shallow NV centers in diamond has potential implications to studies of a wide range of bio-chemical molecules and processes, both \textit{in vitro} and \textit{in vivo}. Our sensing scheme directly detects the magnetic field created by a paramagnetic molecule, without the need for fluorescent tagging. It can be applied to detect and study small molecules and does not suffer from blinking or photo-bleaching~\cite{Giepmans2006}. Together with recent experiments demonstrating NV magnetic sensing of nanoscale ensembles of nuclear spins~\cite{Mamin2013,Staudacher2013}, the present techniques can be used to perform magnetic imaging measurements on single biological molecules, such as proteins, attached to the diamond surface~\cite{Perunicic2013}. Since NV center-based magnetometry was recently shown to be bio-compatible~\cite{LeSage2013}, our approach can be used for \textit{in vivo} magnetic sensing and imaging with enhanced spatial resolution and single-molecule sensitivity. Specifically, our chemical functionalization scheme can be extended to nanodiamonds, functionalizing them to target certain cellular organelles. The population relaxation of NV centers in these nanodiamonds can then be used, for example, to probe ion channel function~\cite{Hall2010a}. Nanodiamonds can also be functionalized with chemical species (spin traps) that react with short-lived free radicals to produce persistent paramagnetic molecules, which can then be magnetically detected using NV centers. Since radicals are thought to play a key role in biochemical processes such as cellular signaling, aging, mutations, and death~\cite{Netzer2009,Halliwell2007,Valko2007,James2004}, the ability to detect small concentrations of short-lived radicals inside living cells with nanometer resolution could be a powerful tool in studying these processes, with possible applications for disease detection and drug development. Radicals also play a crucial role in chemistry, for example in catalysis~\cite{MacMillan2008}.
Finally, when combined with the recently demonstrated scanning probe techniques~\cite{Grinolds2013}, our methods could also find applications in materials science, enabling imaging of rapidly fluctuating magnetic fields near the surfaces of materials such as superconductors~\cite{Kirtley2010}, chiral magnets~\cite{Milde2013}, and topological insulators~\cite{Zhang2009b,Nowack2013}.

We acknowledge Eric Bersin, Yiwen Chu, Mike Grinolds, Nathalie de Leon, Brendan Shields, Joshua Vaughan, Amir Yacoby, and Norman Yao for experimental help and fruitful discussions. This work was supported by the Defense Advanced Research Projects Agency (QuASAR program), NSF, CUA, ARO MURI, Element Six Inc, Packard Foundation, NSERC (NC) and NDSEG (IL).

%\end{document}

%\bibliography{library}


\end{document}